\documentclass[11pt]{article}
\usepackage{a4wide,graphicx}
\usepackage{epsfig,epsf,graphics,psfrag}
\usepackage{times}
\usepackage{float}
\usepackage{color}
\usepackage{amsfonts,amssymb,stmaryrd,latexsym,amsmath}
\usepackage{hhline}

\newcommand{\be}{\begin{equation}}
\newcommand{\ee}{\end{equation}}
\newcommand{\ba}{\begin{eqnarray}}
\newcommand{\ea}{\end{eqnarray}}
\newcommand{\non}{\nonumber}

\newcommand{\al}{&\!\!\!}

\begin{document}
\title{ \hfill{\tiny FZJ-IKP-TH-2010-09, HISKP-TH-10/10}\\[2.5em]
Reconciling the $X(4630)$ with the $Y(4660)$}
\author{Feng-Kun~Guo$^1$\footnote{{\it E-mail address:}
f.k.guo@fz-juelich.de}, Johann~Haidenbauer$^{1,2}$\footnote{{\it E-mail
address:} j.haidenbauer@fz-juelich.de}, Christoph~Hanhart$^{1,2}$\footnote{{\it
E-mail address:} c.hanhart@fz-juelich.de}, and
Ulf-G.~Mei{\ss}ner$^{1,2,3}$\footnote{{\it E-mail address:}
meissner@hiskp.uni-bonn.de}\\[2mm]
  {\it\small$\rm ^1$Institut f\"{u}r Kernphysik and J\"ulich Center for Hadron
  Physics,}\\
          {\it\small Forschungszentrum J\"{u}lich, D--52425 J\"{u}lich, Germany}\\
  {\it\small$\rm ^2$Institute for Advanced Simulation,
          Forschungszentrum J\"{u}lich, D--52425 J\"{u}lich, Germany}\\
  {\it\small$\rm ^3$Helmholtz-Institut f\"ur Strahlen- und Kernphysik and
          Bethe Center for Theoretical Physics,}\\
          {\it\small Universit\"at Bonn, D--53115 Bonn, Germany}}
\date{}

\maketitle
\begin{abstract}
The Belle Collaboration observed an enhancement called $X(4630)$ in the
$\Lambda_c^+\Lambda_c^-$ mass distribution using  initial state radiation. We
demonstrate that the enhancement could be consistent with the $\psi'f_0(980)$
molecular picture of the $Y(4660)$ taking into account the
$\Lambda_c^+\Lambda_c^-$ final state interaction. To test the hypothesis that
the $X(4630)$ and $Y(4660)$ are the same molecular state, we give predictions
for its spin partner, the $\eta_c'f_0(980)$ molecule. High statistic measurements
of the $B$ decays into the $K\Lambda_c^+\Lambda_c^-$ and $K\eta_c'\pi^+\pi^-$
are strongly recommended.
\end{abstract}

\newpage

The recently observed open and hidden charmed hadrons have stimulated many
studies. They challenge our current knowledge of hadron spectroscopy, and
provide us with an opportunity to understand non-perturbative QCD better. Among
these hadrons, the $Y(4660)$ was observed by the Belle Collaboration in the
$\psi'\pi^+\pi^-$ mass distribution using the technique of initial state
radiation (ISR)~\cite{Wang:2007ea}. The mass and width were reported to be
$4664\pm11\pm5$~MeV and $48\pm15\pm3$~MeV, respectively. This structure is very
special because it was neither observed in
$e^+e^-\to\gamma_{ISR}\pi^+\pi^-J/\psi$ \cite{ISRJpsi}, nor in the mass
distributions of a charmed and anti-charmed meson pair in the final states of
electron-positron collisions~\cite{Exp:DDbar,Abe:2007sy}. Furthermore, the
$\pi^+\pi^-$ invariant mass spectrum shows a single peak at the high end, i.e.
towards the mass region of the scalar meson $f_0(980)$. In
Ref.~\cite{Guo:2008zg} it was argued that these facts may be naturally explained
in terms of a hadronic molecular picture, i.e. by $\psi'f_0(980)$ being bound
together in an $S$-wave, while they would challenge other
explanations~\cite{Ding,Qiao,Cotugno:2009ys}.

More recently, the Belle Collaboration reported another structure, called
$X(4630)$, in the $\Lambda_c^+\Lambda_c^-$ invariant mass distribution in
$e^+e^-\to\gamma_{\rm ISR}\Lambda_c^+\Lambda_c^-$~\cite{Pakhlova:2008vn}. The
reported mass is $4634^{+8+5}_{-7-8}$~MeV, and the width is
$92^{+40+10}_{-24-21}$~MeV, consistent with the ones reported for the $Y(4660)$
within two sigma. Based on the tetraquark picture, both structures were proposed to be of the same origin
in Ref.~\cite{Cotugno:2009ys}, however, there is no general consensus on this
issue yet (see e.g. the discussion in the short review~\cite{Godfrey:2009qe}). In
this paper, we shall show that they could also be understood as the same state within
the $\psi'f_0(980)$ hadronic molecular picture, and discuss how this hypothesis
can be tested in future experiments.

In the $\psi'f_0(980)$ hadronic molecular picture, one may expect naively that
the bound state would decay mainly through the decays of the unstable
$f_0(980)$, and hence into the $\psi'\pi\pi$, and the peak in the $\pi\pi$
invariant mass spectrum close to the $f_0(980)$ mass region appears naturally.
While the latter statement is correct, the former one needs to be scrutinized.
The mass of the $Y(4660)$ is higher than open charmed and anti-charmed meson
thresholds, and the $\Lambda_c^+\Lambda_c^-$ threshold.  If the binding energy
$\varepsilon=M_{\psi'}+m_{f_0(980)}-M_{Y(4660)}$ is very small, the coupling of
the bound state to its constituents determined by the
equation~\cite{molecule1,molecule2}
\begin{eqnarray}
\frac{g^2}{4\pi}= 4(M_{\psi'}+m_{f_0(980)})^2\sqrt{{2\varepsilon\over\mu}},
\label{eq:geff}
\end{eqnarray}
with $\mu$ the reduced mass of the $\psi'$ and $f_0(980)$, is small, and so
is the partial width $\Gamma(Y(4660)\to\psi'\pi\pi)$. On the other hand, the
open charm channels have larger phase space, and might have larger partial decay
widths. In fact, there is a well-known example --- the $f_0(980)$ decays mainly
into two pions which have plenty of phase space although it can be understood as
a $K\bar K$ bound state~\cite{molecule2,Weinstein:1990gu}. In this paper, we
shall assume that the $\Lambda_c^+\Lambda_c^-$ is the dominant open charm
channel and study the implications of this assumption. This means we shall
assume the total width of the $Y(4660)$ is given by the sum of the partial
widths into the $\psi'\pi\pi$ and $\Lambda_c^+\Lambda_c^-$, i.e.
\begin{equation}
\Gamma_{Y}^{\rm tot} =
\frac{3}{2}
\Gamma_{Y}^{[\psi'\pi^+\pi^-]}+\Gamma_{Y}^{[\Lambda_c^+\Lambda_c^-]}~,
\end{equation}
where the factor $3/2$ in front of $\Gamma_{Y}^{[\psi'\pi^+\pi^-]}$ is from
isospin symmetry.

The line shape of the $Y(4660)$ is given by its spectral function
\begin{equation}
\rho_{Y}(M)=\frac{M_{Y}\Gamma_{Y}^{\rm tot}(M)}{\left|M^2-M_{Y}^2+\hat \Pi_{Y}(M)
\right|^2} \ ,
 \label{eq:ysf}
\end{equation}
convoluted with phase space, where $M_Y$ is the mass, $\Gamma_{Y}^{\rm tot}(M)$
is the energy-dependent total width, and $\hat \Pi_{Y}(M) = \Pi_{Y}(M)-{\rm
Re}[\Pi_{Y}(M_{Y})]$ is defined as the self-energy with the real part subtracted
at the mass~\cite{achasov}. The self-energy for arbitrary values of $M$ is given
by a dispersion integral (for further details, see Ref.~\cite{Guo:2008zg})
\begin{equation}
\Pi_{Y}(M) = \frac{1}{\pi}\int_{M_{\rm thr}^2}^\infty \!\!\!\!
ds\frac{M_{Y}\Gamma^{\rm tot}_{Y}(\sqrt{s})}{s-M^2-i\epsilon} \ ,
\end{equation}
where $M_{\rm thr}$ denotes the relevant physical threshold. In
Ref.~\cite{Guo:2008zg}, only the decays $Y\to \psi'\pi\pi(K\bar K)$ were
considered. In order to check whether or not the structure observed in the
$\Lambda_c^+\Lambda_c^-$ mass distribution is consistent with the $Y(4660)$
observed in the $\psi'\pi^+\pi^-$, one needs to include the contribution of the
$\Lambda_c^+\Lambda_c^-$ in the total width $\Gamma_{Y}^{\rm tot}$. For that, a
simple Lagrangian for the $Y(4660)\Lambda_c^+\Lambda_c^-$ coupling, which is
assumed to be in an $S$ wave, is used
\be%
\label{eq:YLL}
{\cal L}_{Y\Lambda_c\Lambda_c} = -g_{Y\Lambda_c\Lambda_c} {\bar \Lambda}_c
\gamma^{\mu} Y_{\mu}\Lambda_c~,
\ee%
with $g_{Y\Lambda_c\Lambda_c}$ a dimensionless coupling constant.
Then the cross sections for $e^+e^-\to\gamma_{ISR}\psi'\pi^+\pi^-$ and
$e^+e^-\to\gamma_{ISR}\Lambda_c^+\Lambda_c^-$ are simply given by the
corresponding parts of the spectral function of the $Y(4660)$
\ba%
\sigma(\psi'\pi^+\pi^-) \al=\al N
\frac{M_{Y}\Gamma_{Y}^{[\psi'\pi^+\pi^-]}(M)}{\left|M^2-M_{Y}^2+\hat \Pi_{Y}(M)
\right|^2} , \non\\
\sigma(\Lambda_c^+\Lambda_c^-) \al=\al N
\frac{M_{Y}\Gamma_{Y}^{[\Lambda_c^+\Lambda_c^-]}(M)}{\left|M^2-M_{Y}^2+\hat
\Pi_{Y}(M) \right|^2} , \label{eq:cs}
\ea%
where $\Gamma_{Y}^{[\psi'\pi^+\pi^-]}$ and
$\Gamma_{Y}^{[\Lambda_c^+\Lambda_c^-]}$ are the partial decay widths of the
$Y(4660)$ into the $\psi'\pi^+\pi^-$ and $\Lambda_c^+\Lambda_c^-$ channels, respectively.
The overall normalization constant $N$ is the same for both processes since both
structures were observed by the Belle Collaboration in the ISR processes.

Since the $Y(4660)$ has the quantum numbers $J^{PC} = 1^{--}$, it couples to the
$\Lambda_c^+\Lambda_c^-$ system in an $S$--wave, specifically to the $^3S_1$,
and, therefore, the impact of the final state interaction (FSI) is expected to
be large. In principle, the situation is comparable to $J/\psi$ decays with the
proton--antiproton channel in the final state where FSI effects are known to
play a rather important
role~\cite{Zou:2003zn,Sibirtsev:2004id,Loiseau:2005,Haidenbauer:2006,Entem:2007,Haidenbauer:2008,Chen:2010an}.
Unfortunately, there is no direct experimental information on the interaction
between charmed and anti-charmed baryons. Thus, we have to resort to a model of
the $\Lambda_c^+\Lambda_c^-$ interaction for taking into account FSI effects.
Here we adopt the potential presented in Ref.~\cite{Haidenbauer:2009ad}, which
was derived using SU(4) flavor-symmetry arguments, and compute the Jost function
${\cal J}(M)$ for this interaction. Multiplying the reaction amplitude with the
inverse of the latter quantity, also known as enhancement factor, is practically
equivalent to a treatment within a distorted-wave Born
approximation~\cite{goldberger,Baru:2000}. The width of $Y(4660)\to
\Lambda_c^+\Lambda_c^-$ is then given by
\be%
\Gamma_{Y}^{[\Lambda_c^+\Lambda_c^-]}(M) =
\frac{g_{Y\Lambda_c\Lambda_c}^2}{|{\cal J}(M)|^2}\frac{p}{6\pi}
\left(1+2\frac{M_{\Lambda_c}^2}{M^2}\right) \theta(M-2M_{\Lambda_c}),
\ee%
where $M_{\Lambda_c}$ is the mass of the $\Lambda_c$,
$p=\sqrt{M^2/4-M_{\Lambda_c}^2}$ is its three-momentum in the rest frame of the
$Y(4660)$, and $\theta$ is the step function. In the calculations of
Ref.~\cite{Haidenbauer:2009ad}, the function $1/|{\cal J}(M)|^2$ for the $^3S_1$
channel decreases from about 2 at zero momentum to 0.3 at $p\simeq500$~MeV, and
then slowly approaches unity only at very high momenta. In our calculations, we
parameterize $1/|{\cal J}(M)|^2$ up to $p\simeq500$~MeV with the following
function
\be%
\frac{1}{|{\cal J}(M)|^2} = d \,\frac{p^2+b^2}{p^2+c p+a^2},
\ee%
with the parameter values being $a=247.7$~MeV, $b=1390.4$~MeV, $c=387.3$~MeV,
and $d=0.0677$. Then we set $d=1$, which may always be done because such a
normalization can be absorbed into a redefinition of the coupling constant
$g_{Y\Lambda_c\Lambda_c}$, so that the remaining factor approaches unity
asymptotically, and provides an enhancement to the amplitude close to the
threshold.
\begin{figure}[t]
\begin{center}
\vglue-0mm
\includegraphics[width=0.49\textwidth]{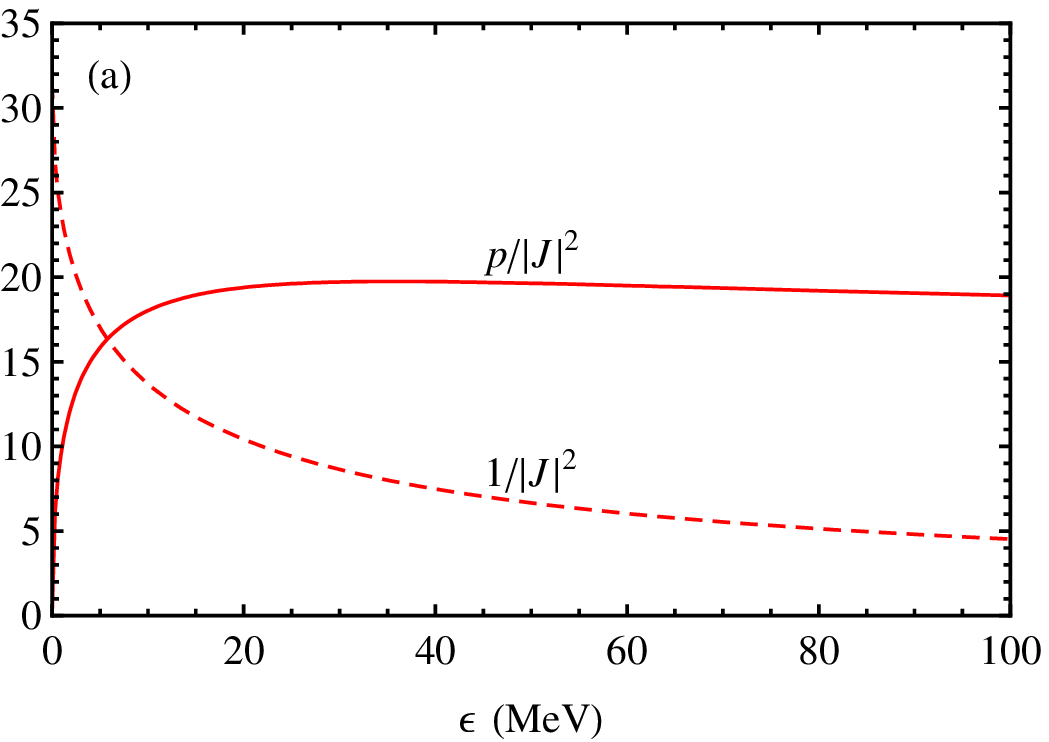}\hfill
\includegraphics[width=0.49\textwidth]{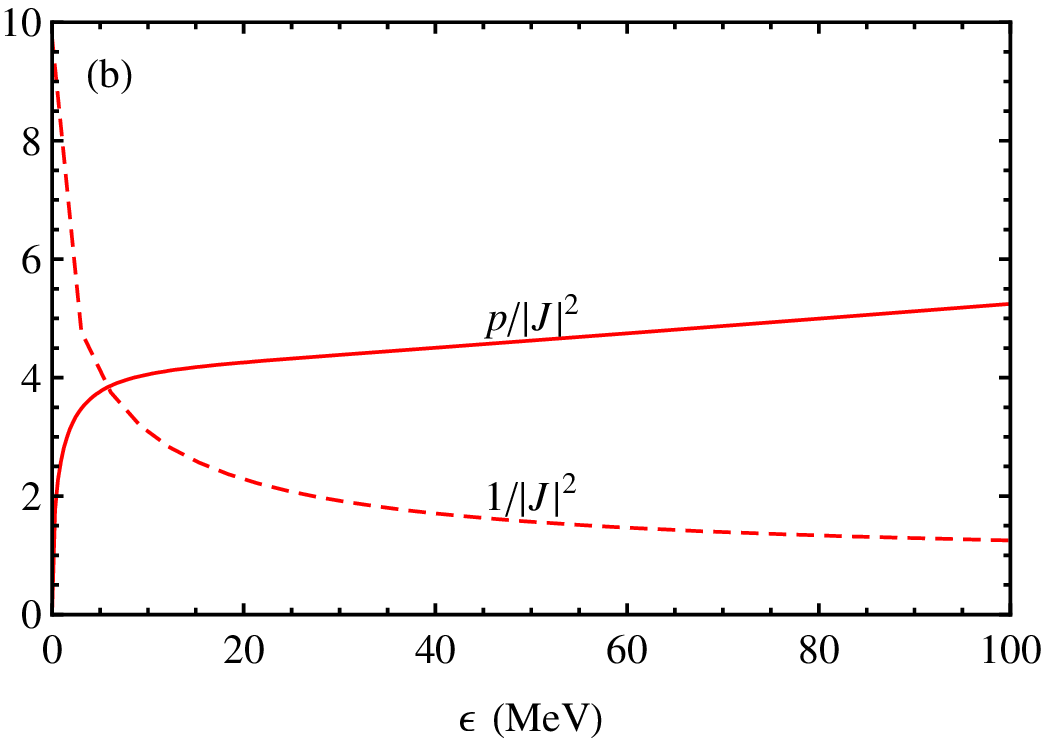}
\vglue-0mm \caption{The FSI enhancement factor
$1/|{\cal J}(M)|^2$ (dashed line) and the quantity
$p/|{\cal J}(M)|^2$ (solid line)
as a function of the excess energy $\epsilon = M-2M_{\Lambda_c}$. The
latter curves are normalized arbitrarily.
(a): the  $^3S_1$ channel; (b): the  $^1S_0$ channel.
\label{fig:fsi}}
\end{center}
\end{figure}
In Fig.~\ref{fig:fsi} (a), the FSI enhancement factor in the $^3S_1$ channel as
well as this factor times the two--body phase space are shown as a function of
the excess energy $\epsilon = M-2M_{\Lambda_c}$.
Note that the central value of the peak observed by the Belle
Collaboration in the $\Lambda_c^+\Lambda_c^-$ mass distribution is about 90~MeV
above threshold, hence it cannot be due to the FSI enhancement solely, as may be
seen from the figure. An opposite claim was made recently in
Ref.~\cite{vanBeveren:2010jz}.

\begin{figure}[t]
\begin{center}
\psfrag{X1}[][]{$M(\psi'\pi^+\pi^-)$~[GeV]}
\psfrag{X2}[][]{$M(\Lambda_c^+\Lambda_c^-)$~[GeV]}
\psfrag{Y1}[][]{$\sigma(\psi'\pi^+\pi^-)$~[pb]}
\psfrag{Y2}[][]{$\sigma(\Lambda_c^+\Lambda_c^-)$~[pb]}
\includegraphics[width=\textwidth]{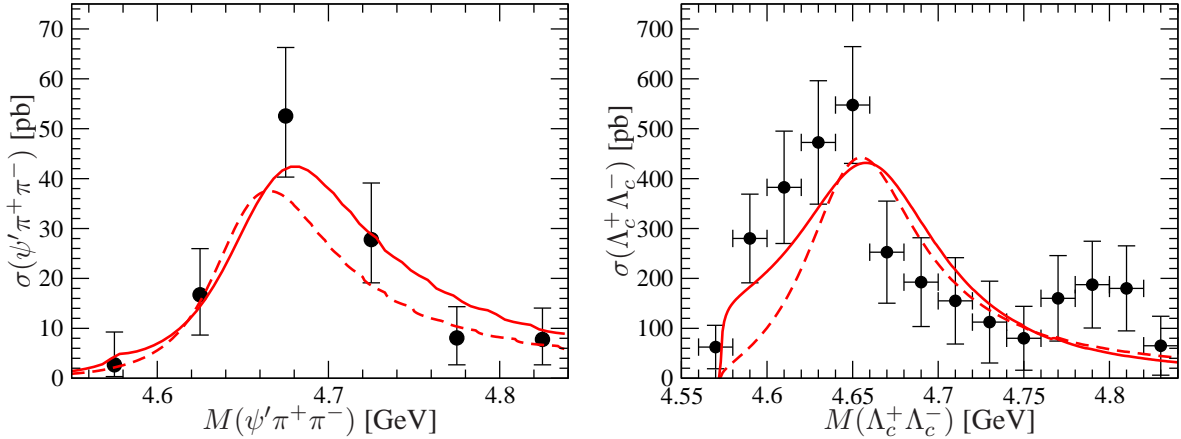}
\vglue-0mm \caption{The $\Lambda_c^+\Lambda_c^-$ and
$\psi'\pi^+\pi^-$ invariant mass spectra. The data are taken from the Belle measurements.
The solid curves are the results of the best fit, and the dashed curves are the results
with FSI effects omitted.
\label{fig:comparison}}
\end{center}
\end{figure}

Using Eqs.~(\ref{eq:cs}), we perform a simultaneous fit to the cross sections of
both processes. For simplicity, we assume that there is no background. Then
there are three free parameters: the normalization constant $N$, the mass of the
$Y(4660)$, $M_Y$, and the $Y(4660)\Lambda_c\Lambda_c$ coupling constant
$g_{Y\Lambda_c\Lambda_c}$. The best fit gives
\be%
N = 237^{+40}_{-36}, \quad M_{Y} = 4662.5^{+0.1}_{-0.2}~{\rm MeV}, \quad
g_{Y\Lambda_c\Lambda_c} = 0.7\pm0.1,
\ee%
with $\chi^2/{\rm d.o.f.} = 1.4$. The uncertainties quoted above are only from the fit,
and do not include an estimate of the systematic uncertainty of the procedure.
In doing the above fit, we chose to use $M_{\psi'}$ as given by the
PDG~\cite{Amsler:2008zz} and the central values of the parameters for the
$f_0(980)$ measured recently by the KLOE Collaboration in the best fit K1 shown
in Table~4 in Ref.~\cite{kloe}, i.e. we used $m_{f_0}=976.8$~MeV,
$g_{f_0K^+K^-}=3.76$~GeV and $g_{f_0\pi^+\pi^-}=-1.43$~GeV. The comparison of
our best fit with the experimental data is presented in Fig.~\ref{fig:comparison},
cf. the solid lines.
Also shown are the results for the case without the $\Lambda_c{\bar \Lambda}_c$ FSI
(dashed lines), which were obtained with the same parameters except for the
coupling constant. We use $g_{Y\Lambda_c\Lambda_c}/|{\cal J}(M_Y)|$ as the
coupling constant for the case without FSI such that it coincides with the FSI modified
coupling at the mass of the $Y(4660)$. From the $\Lambda_c^+\Lambda_c^-$ mass
distribution, one immediately sees the enhancement effect of the FSI on the
cross section close to the threshold. From the best fit, we obtain the partial
widths of the $Y(4660)$
\ba%
\Gamma(Y(4660)\to\psi'\pi^+\pi^-) = 8~{\rm MeV}, \quad
\Gamma(Y(4660)\to\Lambda_c^+\Lambda_c^-) = 93~{\rm MeV},
\ea%
and their ratio is
\be%
\label{eq:yratio} \frac{\Gamma(Y\to
\Lambda_c^+\Lambda_c^-)}{\Gamma(Y\to\psi'\pi^+\pi^-)} = 11.5.
\ee%
The ratio is smaller than the central value $24.8$ extracted in
Ref.~\cite{Cotugno:2009ys,faccini} considering also an interference of the
resonance with a polynomial background. In Ref.~\cite{Cotugno:2009ys} the
authors also treated the $X(4630)$ and the $Y(4660)$ as the same state, however,
in this case as  a compact tetraquark.

At this stage, we want to emphasize that the FSI obtained from the model of
Ref.~\cite{Haidenbauer:2009ad} is afflicted with sizeable
uncertainties. However, it incorporates all essential features one expects
from a realistic FSI, specifically it is generated by solving a 
scattering equation and it includes effects from the presence of annihilation
channels. Therefore, it should be sufficient to give an illustration for the
FSI effect in the problem at hand. The $\Lambda_c^+\Lambda_c^-$ interaction of
Ref.~\cite{Haidenbauer:2009ad} contains two parts --- an elastic part based on
meson exchange and derived via SU(4) flavor symmetry, and an optical potential
to simulate annihilation processes. In order to check in-how-far changes in
the FSI influence our results we varied the strength of the optical potential
by factors in the range from 1/2 to 2. It turned out that these variations
only have a marginal effect on the resulting invariant mass distributions from
the best fit.

It should be clear that what we discussed above is only a possible scenario. 
The fact that one can obtain a combined
 fit of the $\Lambda_c^+\Lambda_c^-$ and the $\pi\pi\psi'$ channels also
in the molecular picture does not
prove that the $X(4630)$ and the $Y(4660)$ are the same state. Observables
should be found to further support or disprove this hypothesis. In
this context, it is important to investigate the spin partner. Heavy quark spin
symmetry in any case predicts the existence of a spin partner, however, the
scenario outlined implies some very specific properties of that spin partner
with respect to its mass and decay properties, as we will discuss now.

In Ref.~\cite{Guo:2009id}, based on heavy quark spin symmetry, we predicted the
presence of an $\eta_c'f_0(980)$ bound state, called $Y_\eta$, as the spin
multiplet partner of the $\psi'f_0$ bound state. The mass of the $Y_\eta$ should
satisfy
\be%
M_{Y_\eta} = M_{Y(4660)}-(M_{\psi'}-M_{\eta_c'}) \label{eq:myeta}
\ee%
to a high precision. Using the best fit value for the $Y(4660)$ mass given above
and $M_{\eta_c'}=3637\pm4$~MeV~\cite{Amsler:2008zz}, one gets
$M_{Y_\eta}=4613\pm4$~MeV where the uncertainty is dominated by the one from the
$\eta_c'$ mass. Based on the same formalism as above, the line shape of the
$Y_\eta$ in the $\eta_c'\pi^+\pi^-$ and the $\Lambda_c^+\Lambda_c^-$ may be
predicted. Heavy quark spin symmetry indicates that the coupling of the $Y_\eta$
to the $\Lambda_c^+\Lambda_c^-$ has the form, cf. Eq.~(\ref{eq:YLL}),
\be%
{\cal L}_{Y_\eta\Lambda_c\Lambda_c} = ig_{Y\Lambda_c\Lambda_c} {\bar \Lambda}_c
\gamma^{5} Y_\eta \Lambda_c ,
\ee%
with the same coupling constant as the $Y(4660)$.

\begin{figure}[t]
\begin{center}
\psfrag{X1}[][]{$M(\eta_c'\pi^+\pi^-)$~[GeV]}
\psfrag{X2}[][]{$M(\Lambda_c^+\Lambda_c^-)$~[GeV]}
\psfrag{Y}[][]{Line shape [arbitrary units]}
\includegraphics[width=\textwidth]{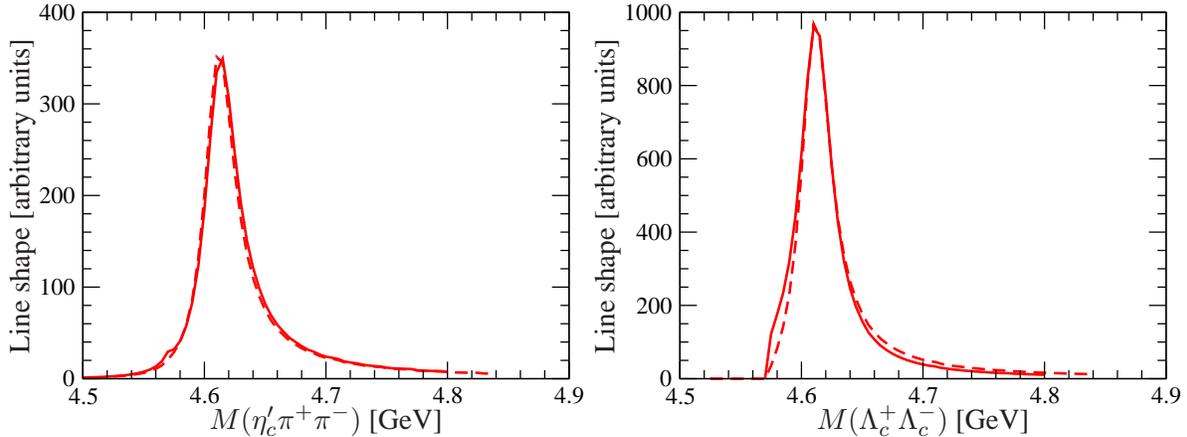}
\vglue-0mm \caption{Predictions of the $Y_\eta$
line shapes in the $\eta'\pi^+\pi^-$ and $\Lambda_c^+\Lambda_c^-$ in arbitrary units.
The solid and dashed curves represent results with and without FSI, respectively.
\label{fig:yeta}}
\end{center}
\end{figure}
In Fig.~\ref{fig:yeta}, the predictions for the $Y_\eta$ line shapes in the
$\eta'\pi^+\pi^-$ and $\Lambda_c^+\Lambda_c^-$ channels are shown in arbitrary
units, however, with the relative normalization fixed. With the FSI, now in the $^1S_0$ partial wave and calculated again from
the $\Lambda_c^+\Lambda_c^-$ model of Ref.~\cite{Haidenbauer:2009ad}, shown in
Fig.~\ref{fig:fsi} (b), the predicted line shapes are given by the solid curves,
while the ones without FSI are given by the dashed curves. The $Y_\eta$ mass is
only about 40~MeV higher than the $\Lambda_c^+\Lambda_c^-$ threshold, as a
result the width of the $Y_\eta$ is much smaller than that of the $Y(4660)$, and
thus the line shapes are much narrower. The partial widths for decay into the
$\eta_c'\pi^+\pi^-$ and the $\Lambda_c^+\Lambda_c^-$ channels are 8~MeV and
22~MeV, respectively. The ratio
\be%
\frac{\Gamma(Y_\eta\to
\Lambda_c^+\Lambda_c^-)}{\Gamma(Y_\eta\to\psi'\pi^+\pi^-)} = 2.7
\label{yetaratio}
\ee%
is much smaller than the one for the $Y(4660)$ as a result of smaller phase
spaces. Furthermore, the effect of the FSI is not so significant anymore. We
expect that within other models for the spin partner of the $Y(4660)$ the
discussed properties, especially the mass and the ratio of
Eq.~(\ref{yetaratio}), will be very different.

In summary, taking into account the $\Lambda_c^+\Lambda_c^-$ FSI, we found that
the $X(4630)$ may be described as the same state as the $Y(4660)$ in the
$\psi'f_0(980)$ bound state picture. One notices that there should be other open
charm decay channels, such as decays into charmed and anti-charmed mesons. We
checked that an additional constant width from other possible decay channels of
less than 30~MeV may still be accommodated. In principle, a polynomial
background as in Ref.~\cite{Cotugno:2009ys} allows one to improve the fit. Also
possible interferences of the $X(4630)$ or $Y(4660)$ with other resonances, such
as highly excited $\psi$ resonances, could have an impact on the analysis.
However, neither of these effects is under control quantitatively given the
current quality of the experimental data.
 Hence, in our analysis, we refrain from
considering them to reduce the number of parameters. Within the molecular
picture for the $Y(4660)$, the presence of a $Y_\eta$ with a mass given by
Eq.~(\ref{eq:myeta}) as the spin partner of the $\psi'f_0(980)$ bound state is
almost unavoidable, since the spin-dependent interactions are highly suppressed
by $1/m_c^2$, with $m_c$  the charm quark mass~\cite{Guo:2009id}. Other
models of the $Y(4660)$ should also provide a spin partner, but most probably
with a different mass and different decay patterns. Thus, in order to test the
molecular picture it is important to search for the $Y_\eta$ experimentally, for
instance in the decays $B^{\pm}\to\eta_c'K^{\pm}\pi^+\pi^-$ which is expected to
have a large branching fraction~\cite{Guo:2009id}.

At last, we want to mention that a related observation was made by the BaBar
Collaboration in the reaction
$B^-\to\Lambda_c^+\Lambda_c^-K^-$~\cite{Aubert:2007eb}. They observed a
structure at $2931\pm3\pm5$~MeV in the $\Lambda_c^+K^-$ mass distribution. In
the paper, the $\Lambda_c^+\Lambda_c^-$ mass distribution is also provided,
where one can see clearly two peaks. The measured branching ratio of the decay
$B^-\to\Lambda_c^+\Lambda_c^-K^-$ is of order $10^{-3}$~\cite{Aubert:2007eb},
which is several orders higher than the naive expectation $10^{-8}$ since this
three-body decay is color-suppressed and with a small phase
space~\cite{Cheng:2005vd}. In Ref.~\cite{Cheng:2005vd} Cheng {\it et al.} showed
that the high suppression could be diminished, if there was a narrow hidden charm
state with a mass of order $4.6-4.7$~GeV or a charmed baryon, which was assumed
to have $J^P=1/2^+$, coupled to the $\Lambda_c^+K^-$. We notice that the
positions of the double peaks coincide with the masses of the $Y(4660)$ and the
predicted $Y_\eta$. However, they could also be due to a charmed baryon $\Xi_c$
with $J^P=3/2^+$ --- we found that a  $J^P=1/2^+$ $\Xi_c$ baryon, as used in
Ref.~\cite{Cheng:2005vd}, cannot describe the double peak structure in the
$\Lambda_c^+\Lambda_c^-$ mass distribution.
Also some interference of a charmed baryon with the charmonia is possible. 
 Better data with higher statistics,
especially better Dalitz plots, would be very helpful in illuminating the
situation.

\vspace{5mm}
We would like to thank R. Faccini and A. Polosa for communications concerning
their work. This work is partially supported by the Helm\-holtz Association
through funds provided to the virtual institute ``Spin and strong QCD''
(VH-VI-231) and by the DFG (SFB/TR 16, ``Subnuclear Structure of Matter''). We
also acknowledge the support of the European Community-Research Infrastructure
Integrating Activity ``Study of Strongly Interacting Matter'' (acronym
HadronPhysics2, Grant Agreement n. 227431) under the Seventh Framework Programme
of EU.

\end{document}